\documentclass[preprint]{aastex}

\shorttitle{X-rays from an O-star Population}
\shortauthors{Helfand and Moran}

\begin{document}

\title{The Hard X-ray Luminosity of OB Star Populations: Implications for the
Contribution of Star Formation to the Cosmic X-ray Background}

\author{David J.\ Helfand}
\affil{Columbia University, Dept.\ of Astronomy, 550 West 120th Street,
       New York, NY~10027} 
\email{djh@astro.columbia.edu}

\author{Edward C.\ Moran\altaffilmark{1}}
\affil{University of California, Dept.\ of Astronomy, 601 Campbell Hall,
       Berkeley, CA 94720-3411}
\email{edhead@jester.berkeley.edu}

\altaffiltext{1}{{\it Chandra\/} Fellow.}

\begin{abstract}

We present an empirical analysis of the integrated X-ray luminosity arising
from populations of OB stars. In particular, we utilize results from the
All-Sky Monitor on {\sl RXTE\/}, along with archival data from previous
missions, to assess the mean integrated output of X-rays in the 2--10 keV
band from accreting
early-type binaries within 3 kpc of the Sun. Using a recent OB star census
of the Solar neighborhood, we then calculate the specific X-ray luminosity
per O star from accretion-powered systems. We also assess the contribution to 
the total X-ray luminosity of an OB population from associated T Tauri stars,
stellar winds, and supernovae. We repeat this exercise for the major
Local Group galaxies, concluding that the total X-ray luminosity per
O star spans a broad range from 2 to $20 \times 10^{34}$ erg s$^{-1}$. Contrary
to previous results, we do not find a consistent trend with metallicity; in
fact, the specific luminosities for M31 and the SMC are equal, despite having
metallicities which differ by an order of magnitude. In light of these
results, we assess the fraction of the observed 2--10 keV emission from
starburst galaxies that arises directly from their OB star populations,
concluding that,
while binaries can explain most of the hard X-ray emission in many local
starbursts, a significant additional component or components must
be present in some systems. A discussion of the nature of this additional
emission, along with
its implications for the contribution of starbursts to the cosmic X-ray
background, concludes our report.
\end{abstract}


\section{Introduction}

Early-type stars are themselves modest X-ray emitters, with a mean output in
the 0.1--10 keV band of only $\sim10^{-7}$ of their optical/UV flux. Thus,
even a population of a million OB stars, typical of that found in a galaxy
undergoing a vigorous starburst, will only produce an X-ray luminosity of
$\sim 10^{38}\> {\rm ergs\> s^{-1}}$. A single young neutron star similar to
that in the Crab Nebula or a single O-star binary with a neutron star or
black hole companion will thus
outshine the entire population of main sequence stars. In order to determine
the expected contribution of a young stellar population to the X-ray 
luminosity of a galaxy, then, it is necessary to estimate accurately the
specific X-ray luminosity per O star, most of which comes from the deceased
segment of the population. Such an exercise is of interest in light of the
relatively high X-ray luminosities of starburst galaxies and the potential
contribution of such objects to the cosmic X-ray background. 

We attempt here a systematic, empirical
census of the direct contributions OB stars, their lighter siblings, and their
stellar remnants make to the hard (2--10 keV) X-ray luminosity of a starburst
galaxy by calculating the specific X-ray luminosity per O-star in the
Solar neighborhood, the Galaxy as a whole, and the other members of the Local
Group. We begin (\S~2) with a cautionary tale concerning the calculation of
long-term, mean X-ray luminosities for high-mass X-ray binaries by surveying
the literature on the most luminous such system in the Local Group, SMC X-1.
In \S~3, we compile a list of all high-mass, accretion-powered X-ray binaries
that lie within 3 kpc of the Sun and, using the Rossi X-ray Timing Explorer ({\sl RXTE}) All-Sky 
Monitor database and other archival data, compute the total X-ray luminosity
within this volume arising from such systems. Using a recent census of OB stars
in the solar vicinity, we then calculate the specific X-ray luminosity per
O star from accreting systems. This local estimate is then compared to that
for the Galaxy as a whole. The following section (\S~4) repeats this analysis
for the other Local Group galaxies, concluding with a commentary on the
reported dependence of this value on metallicity. In \S~5, we consider
the other contributions an OB star population makes to the integrated
X-ray luminosity of a galaxy---from associated T Tauri stars, stellar
winds, and supernovae. We then go on (\S 6) to assess the fractional contribution
these direct sources of X-ray emission make to the total hard X-ray
luminosity of starburst galaxies. We conclude with
a summary of our results, a brief discussion of additional possible
contributions to the X-ray luminosity of starburst galaxies, and the
implications of these results for the origin of the cosmic X-ray background.

\section{The Mean Observed X-ray Luminosity of SMC X-1}

The High-Mass X-ray Binary (HMXB) SMC X-1, the only persistent, bright,
accretion-powered source in the Small Magellanic Cloud (SMC), consists of
a B0 supergiant primary accompanied by a neutron star with a 0.7 s pulse period
in a 3.9 d orbit. The system is the most luminous X-ray binary in the Local
Group, and is frequently cited as ``superluminous,'' given that its nominal
X-ray luminosity exceeds the Eddington limit for a $1.4\> M_{\odot}$ neutron
star.  But just what is the mean, integrated luminosity of SMC X-1 as
observed from Earth?

The most complete recent catalog of X-ray binaries is that of van Paradijs
(1995) which lists, among other system parameters, a maximum and minimum (if
available) reported flux density for each source. In the case of SMC X-1, these
values are 57 and 0.5 $\mu$Jy. The incautious reader might adopt either the
maximum value (especially for the majority of sources in the catalog for
which only one value is given), or simply
average the two numbers to estimate the mean X-ray luminosity of the source. In
fact, to convert these values to a mean observed X-ray luminosity requires
adoption of a distance to the SMC,
a mean spectral form for the X-ray emission, and a bandwidth over which the
emission is integrated, as well as a description of the temporal behavior of
the source. Compiling the variety of assumptions actually adopted in
the literature is instructive: 45 kpc $< D <$ 70 kpc (Howarth 1982; Seward
\& Mitchell 1981);
$0.9 < {\Gamma} < 3.28$; (Angelini, White, \& Stella 1991; Coe et al.\ 1981); $3 \times 10^{20}<
N_{\rm H}< 3.6 \times 10^{22}$ cm$^{-2}$ (Kahabka \& Pietsch 1996; Davison 
1977); and 0.2--2.4 keV to 2--100 keV 
(Kahabka \& Pietsch 1996; Coe et al.\ 1981). As a consequence, quoted luminosities range from $6 
\times 10^{36}$ erg $\rm s^{-1}$ (Seward \& Mitchell 1981) to $3 \times
10^{39}$ ergs $\rm s^{-1}$ (Price et al.\ 1971); far from all of this uncertainty results from the
source's intrinsic variability. Furthermore, some reports undertake systematic
data editing that bias the flux estimates upward---leaving out data during the
16\% of the time the X-ray source is in eclipse, ignoring periods when the
source is at an undetectable level for a given instrument, etc.---which, while
usually well-documented and appropriate for the task at hand, require reversal
when attempting to define the source's mean contribution to its galaxy's X-ray 
luminosity.

To continue with this example, we adopt a distance to the Small Cloud of 65 kpc
and, for SMC X-1 itself, we employ spectral parameters $\Gamma= 1.1$,
$N_{\rm H}=6 \times 10^{20}$ (Angelini et al.\ 1991) and a cutoff energy of
6.5 keV; while these parameters ignore a reported soft ({\it kT} $\sim0.25$ keV)
component, they suffice to illustrate our point. Correcting the mean fluxes
for SMC X-1 reported over monitoring times of weeks to a decade by
Angelini et al.\ (1991), Wojdowski et al.\ (1998), Whitlock \& Lochner (1994),
Gruber \& Rothschild (1984), and Levine et al.\ (1996) to the 2--10 keV band,
we find {\it all\/} are consistent within two sigma with the value of
$2.2 \times 10^{38}$ ergs $\rm s^{-1}$, or 17 $\mu Jy$, a factor of
5 below the highest value in the literature, and a factor of 3.5 below the
cataloged flux. This value is below the Eddington limit for a neutron star
mass of $1.9\> M_\odot$ even before correcting for the lower metallicity
of the accreting material in the SMC; more importantly, however, it represents 
the appropriate value to adopt in any summation of the integrated X-ray
luminosity of the SMC OB-star population (see \S~4.2) and illustrates the
need for caution when undertaking such a task.

\section{The X-ray Luminosity per O Star for Accreting Systems in the
Milky~Way}

\subsection{The Solar Neighborhood}

We begin by examining in detail the populations of HMXBs and OB stars where
our information is most complete---within 3 kpc of the Sun. In Table
1, we list all 57 accretion-powered X-ray binaries with high-mass stellar
primaries ever reported in the literature as lying within this
distance. Much of the data are taken from the catalog of van Paradijs (1995);
a survey of the literature in the intervening five years has been used to
bring the list up to date. The first column contains the original source
name, followed by a vernacular name (if any). Columns 2--4 list the optical
counterpart name and J2000 coordinates. These have been taken from the
Hipparcos/Tycho catalogs (Perryman et al.\ 1997; H\o g et al.\ 1997) if
available (see also Chevalier \& Ilovaisky 1998), and
elsewise from the Guide Star Catalog, or from the highest precision position
reported in the literature. For sources without confirmed optical
identifications, the best available X-ray coordinates are given, and the star
listed is the brightest object in the error circle. The source
of the position is given in column 5; if the positional uncertainty is
greater than $10''$ and/or the identification is uncertain,
its value is given in parentheses. Columns 6 and 7
give the star's visual magnitude and spectral type; these are taken
from the Hipparcos or Tycho catalogs when available, and otherwise from the literature. The source distances follow (cols. 8--10);
they include the smallest distance reported in the literature, a best estimate
(based in a few cases on Hipparcos parallaxes, but mostly on a qualitative
assessment of the literature), and the maximum plausible distance. For stars
with Hipparcos observations, the 1.5~$\sigma$ lower limit is given. If a
distance upper limit is greater than 3.0 kpc, it is quoted as ``$>3$~kpc'',
and the source's contribution to the local X-ray binary luminosity is
computed as if
it were at 3 kpc for purposes of calculating an upper limit to this value.
Clearly, if the source lies at a greater distance, its inferred luminosity
would be higher, but its contribution to the quantity of interest is zero,
making this a conservative approach to calculating an upper limit to the X-ray
luminosity of the population as a whole.

The remaining columns of the Table report source fluxes and luminosity
measurements. When
spectral parameters are provided in the literature, they have been used to
correct the observed flux
to the 2--10 keV band. When only an instrumental flux and bandwidth are quoted,
we have adopted a power-law spectral form with a photon index of
1.0 and a plausible column density for the adopted distance(s) (using
$3 \times 10^{21}\> {\rm cm^{-2}\> kpc^{-1}}$ unless excess extinction is
indicated). Our results are not sensitive to this specific choice of parameters:
varying the power law index over the range $0 < \Gamma < 3$ and the column
density from $0.1< N_H<10 \times 10^{21}$ cm$^{-2}$ changes the inferred
luminosities by $<20\%$.

Since our goal is to calculate the best available long-term mean integrated
flux for each source, we have utilized the {\sl RXTE\/} All-Sky Monitor
(ASM; Levine et al.\ 1996) light curves when available; more than half the sources have
such light curves with largely continuous coverage (80\% to 98\%) over more than 1650 days.
For each of these, we include the number of days in the light curve for
which no flux is available, the global mean flux, the mean flux plus 2~$\sigma$
(since many sources are not detected on most days, this value is useful as
a 2~$\sigma$ upper limit), the number of days the source flux exceeded 4~$\sigma$,
the mean flux on those days, and the first and tenth brightest daily
mean fluxes in the 4.5-year interval; the latter values are included in order
to ascertain whether or not one, or a few, large outbursts dominate the
time-integrated luminosity. The 26 sources not included in the
ASM database have never been significantly above the ASM threshold at any time
throughout the last 4.5 years (R. Remillard, private
communication). For these sources, we adopt an upper limit of 1.0 ASM ct ${\rm s^{-1}}$
and calculate the luminosity limit for each source as described above.

This limit is very conservative. As shown below, the mean ASM flux value for
a truly absent source (SMC X-3, for example) over the ASM monitoring interval is
$\sim 0.03$ ct s$^{-1}$. The fact that all of the 31 regularly monitored sources
have mean values a factor of five or more above this limit suggests that they
are often present just below the detection threshold. But there is no such
evidence for those sources which have never crossed the ASM detection threshold,
suggesting that a reasonable upper limit to their contribution could be
at least an order of magnitude lower.

In addition, for all sources, we have searched the High Energy Astrophysics 
Science Archive Research Center (HEASARC)
X-ray Binary Catalog which archives all observations of X-ray binaries in
the Center's large collection of databases. For each source, we list the
number of detections (col. 18), the maximum count rate and, in column 21,
the catalog from which the count rate was taken. In no case
does the integrated luminosity in a major outburst exceed the
integrated luminosity over 30 years derived from the ASM 4.5-year averages
or our conservative upper limits thereto.

As can be seen from the final line in Table 1, our best estimate for
the integrated, mean 2--10 keV
luminosity of the HMXB population within 3 kpc of the Sun is $\sim2 \times 10^{37}\> {\rm ergs\> s^{-1}}$.
Roughly one-third of this total comes from the black hole system Cyg X-1,
one third from a handful of neutron star binaries such as Vela X-1 and
4U1700--37, and the final third from the upper limits adopted for the
26 sources not detected in the ASM. The estimate is conservative, since
it includes all the sources not detected in the ASM as contributing at
1 ct s$^{-1}$. Furthermore, nine of the systems (contributing
11\% of the total flux) have nominal distances beyond 3.0 kpc, but are included
because their distance uncertainties allow membership in our volume-limited
sample.

If we take the 2~$\sigma$ upper limits for {\it all\/} objects, the integrated value only increases by 40\%. Adopting the additional extreme assumption that
all sources are at their maximum allowed distances still does not
raise the conservative best estimate by a factor of two. While examination
of the 30-year history of all sources does show much higher luminosities in
some cases for brief intervals, there is no evidence to suggest that the
last 4.5 years of ASM data is in any way atypical. Thus, we conclude that
the HMXB population within the 38 kpc$^3$ volume surrounding the Sun produces
a 2--10 keV X-ray luminosity of 2--3 $\times10^{37}\> {\rm ergs\> s^{-1}}$.

The final step in calculating the specific X-ray luminosity per O-star in
the solar neighborhood is to find the number of O-stars within 3.0 kpc. We
use the recent (unpublished) compilation of K. Garmany (private communication).
She finds a total of 351 spectroscopically confirmed stars of types O3 through
O9 out to 1.95 kpc from the Sun; the number in bins of constant projected area
(excepting a local minimum) is roughly constant out to this distance, suggesting
incompleteness is not a problem. In addition, there are 1915 B0--B2 stars,
plus a total of 772 stars with the colors of OB stars which lack spectroscopic
types. Adopting the same
O/B ratio as for the classified stars (18\%) suggests as many as $\sim$ 140
additional O stars should be added to the total. Extrapolating with a constant
surface
density out to 3.0 kpc, then, yields a total of 1165 O stars within the volume.
The specific accretion-powered X-ray luminosity per O star is 
$1.6 \times 10^{34}\> {\rm ergs\> s^{-1}\> star^{-1}}$, with a conservative upper
limit (2~$\sigma$ X-ray source upper limits, maximum distances, and no O stars
in the unclassified portion of the stellar sample) of $3.4 \times 10^{34}\>
{\rm ergs\> s^{-1}\> star^{-1}}$.

The census conducted here allows an estimate of the fraction of early
type stars that eventually form X-ray binary systems. For example, there are
3038 stars of types O3 through B2 in the Garmany compilation, implying a
total of $\sim8700$ stars within the 3 kpc distance (correcting for a modest
incompleteness evident in the B0--B2 star counts); in this same region
(using best-estimate distances), we
currently know of 24 HMXBs with primaries of these types, as well as 22 other
accreting systems for which the spectral class is too poorly established to
include them unambiguously. Thus, $\sim0.1-0.2\%$ of all the early-type stars
are currently active accretion-powered X-ray sources. Since the HMXB phase
lasts for a few percent of an OB star's lifetime (Portegies-Zwart \& Verbunt
1996), $\sim2-5\%$ of all OB stars must produce an HMXB. This is roughly
consistent with population synthesis studies (Dewey \& Cordes 1987; Meurs \&
van den Heuvel 1989; Dalton \& Sarazin 1995; Lipunov, Postnov, \& Prokhorov 1997; Terman, Taam, \& Savage 1998; Portegies-Zwart \& van den Heuvel 1999), although the
predictions of such calculations are quite sensitive to the assumed kick
velocity imparted to neutron stars at birth. In our limited sample, at least,
the fraction of X-ray active O stars is similar to that for B stars; given their
shorter lifetimes, the fraction of HMXBs produced must be larger, consistent
with the notion that kick velocities become increasingly sucessful at
unbinding binaries as the mass of the companion star decreases. Note that 
these statistics include HMXBs with luminosities as low as $\sim 10^{33}$ erg 
s$^{-1}$; the fraction of systems with persistent luminosities $>5 \times 
10^{35}$ erg s$^{-1}$ is an order of magnitude smaller.

\subsection{The Whole Galaxy}

The integrated Lyman continuum luminosity of the Milky Way is  $\sim 2.1
\times 10^{53}\> {\rm photons\> s^{-1}}$ (van den Bergh \& Tammann 1991). 
Using Table 5 of Vacca (1994)
for solar metallicity, a Salpeter mass-function slope of 2.35, and a
mass upper limit of 80~$M_{\odot}$ implies a total Galactic population of O stars
of $\sim22,300$. This is $\sim 20\%$ greater than a straightforward
extrapolation from the local population discussed above to the full Galactic
disk (R = 12 kpc), consistent with the observed enhancement of star formation
activity in the inner Galaxy. It is also consistent with the claim of
Ratnatunga and van den Bergh (1989) that the total Pop I content of the Galaxy
is $1000 \pm250$ times that found in a 1 kpc$^2$ area of the disk centered on the Sun, and with an estimate (van den Bergh \& Tammann 1991) based on
counts of embedded O stars from IRAS observations (Wood \& Churchwell 1989).
The uncertainty in the number of O stars is probably less than 50\%.

The total number of HMXBs in the Galaxy is less well constrained. Eight
persistent sources are known with luminosities greater than $10^{36.5}\>
{\rm ergs\> s^{-1}}$ (see Dalton \& Sarazin 1995); these produce a total
{\it peak\/} X-ray luminosity (see \S~2) of $\sim 1.9 \times 10^{38}\>
{\rm ergs\> s^{-1}}$. Other bright, unidentified X-ray sources in the
Galactic plane could add to this population; Dalton \& Sarazin's population
synthesis model predicts 12 sources with $L_{\rm x}>10^{37}\> {\rm ergs\>
s^{-1}}$ and 43 sources with $L_{\rm x}>10^{36}\> {\rm ergs\> s^{-1}}$.
Tripling the four known sources with $L_{\rm x}>10^{37}\> {\rm ergs\>
s^{-1}}$ to match this prediction would yield a luminosity contribution
of $5.3 \times10^{38}\> {\rm ergs\> s^{-1}}$. Integrating the Dalton and Sarazin
predicted luminosity function down to $10^{33}\> {\rm ergs\> s^{-1}}$ and adding
the Be star population at a mean luminosity of $10^{33.5}\> {\rm ergs\>
s^{-1}}$ yields a nominal HMXB luminosity for the Galaxy of $8 \times10^{38}\>
{\rm ergs\> s^{-1}}$.  This value is dominated by the luminous sources,
as is observed to be the case in the solar neighorhood. Adding the
predicted flux from sources with  $L_{\rm x}<10^{36.5}\> {\rm ergs\> s^{-1}}$
to the observed $L_{\rm x}$ of the eight bright sources yields a lower limit of $3.5
\times10^{38}\> {\rm ergs\> s^{-1}}$ for the Galaxy's total $L_{\rm x}$.
Dividing the nominal value by the O star population derived above gives
$3.5 \times 10^{34}\> {\rm ergs\> s^{-1}\> star^{-1}}$, the same as
the upper limit on locally derived value and, again, uncertain by a factor $\sim2$.

\section{The X-ray Luminosity per O Star in Local Group Galaxies}
 
The specific luminosities derived above depend on a number of factors which
could well be different in environments such as the nuclear starbursts to
which we ultimately wish to apply our results. For example, metallicity
can introduce a variety of effects: lower metallicity (1) increases the
Eddington luminosity of accreting sources by decreasing the X-ray scattering 
cross section of the infalling material, (2) lowers the mass of a star of a given 
spectral type (and thus changes the conversion factor between the number of Lyman 
continuum photons and the number of O stars), and, possibly, (3) changes the
ratio of black holes to neutron stars formed in stellar collapse (Hutchings
1984; Helfand 1984). In order to explore the range of specific X-ray luminosities in different galactic environments, we have repeated the exercise
of counting HMXBs and O stars in the four largest external members of the Local Group.

\subsection {The LMC}

The first X-ray sources discovered in the Large Magellanic Cloud were detected
with non-imaging rocket-borne instruments thirty years ago (Price et al.\ 1971). Since then,
systematic imaging surveys have been carried out by the {\sl Einstein\/}
Observatory (Long, Helfand, \& Grabelsky 1981; Wang et al.\ 1991),
{\sl EXOSAT\/} (Pakull et al.\ 1995; Pietsch et al.\ 1989), and {\sl ROSAT\/}
(see Haberl \& Pietsch 1999, although the complete results have yet to
be published).  In addition, ten X-ray binary candidates have been monitored
by the {\sl RXTE\/} ASM, allowing us to calculate accurate mean fluxes
on timescales of years.

Of the four bright, persistent accreting binaries in the LMC, one (LMC X-2)
is a low-mass system and does not concern us here. LMC X-1 and LMC X-3 are
both strong black hole candidates (Hutchings et al.\ 1987; Cowley et al.\ 1983) with steep X-ray spectra
($\Gamma \approx 3-4$; White \& Marshall 1984); LMC X-4 is a HMXB pulsar with a flat power-law
index of $\Gamma\approx 0.7$ (Kelley et al.\ 1983). Using these spectral parameters and the ASM
mean count rates calculated as above, the integrated 2--10 keV luminosity of
these three sources is $2.7 \times 10^{38}\> {\rm ergs\> s^{-1}}$ for an
adopted distance of 50 kpc.

In addition to these persistent sources, the original imaging surveys,
high resolution images of 30 Doradus (Wang \& Helfand 1991; Wang 1995), studies of variable sources
in the {\sl ROSAT\/} data (Haberl \& Pietsch 1999), and monitoring observations
by {\sl Ariel V}, {\sl RXTE}, {\sl CGRO}, etc. have led to the detection of
an additional fourteen HMXB candidates in the Large Cloud. Several of these
have been confirmed as Be-pulsar systems through the detection of X-ray
pulses, although the majority have neither firm optical or X-ray confirmation
of their identity. Seven have been monitored by the ASM for periods ranging
from $\sim800$ to $1650$ days. While most of these have been detected
on a few occasions, none has a mean flux in excess of 0.15 ASM ct~s$^{-1}$;
for nominal spectral parameters of $\Gamma=1$ and $N_{\rm H}\approx 1 \times
10^{21}\> {\rm cm^{-2}}$, this corresponds to an upper limit of $1.4\times
10^{37}\> {\rm ergs\>  s^{-1}}$. Of the remaining non-ASM sources, none has ever
been reported above a luminosity of $1\times 10^{36}\> {\rm ergs\> s^{-1}}$
for more than a single day in outburst. Finally, there remain several
dozen unidentified X-ray point sources from the {\sl Einstein\/} survey.
Although the majority of these are background interlopers, some could be
additional HMXBs;  however, the integrated luminosity of the brightest ten
sources is
$<10^{37}\> {\rm ergs\> s^{-1}}$. An even larger number of (mostly fainter)
point sources are to be found in the {\sl ROSAT\/} survey, but, again, the
majority will be interlopers, and the integrated luminosity of any LMC HMXBs
will not affect our sums by more than a few percent. Thus, we estimate the
total accretion luminosity of the OB population in the LMC to be 
$\sim3\times 10^{38}\> {\rm ergs\> s^{-1}}$, or roughly half that of the
Milky Way.

Kennicutt \& Hodge (1986) have derived the total Lyman continuum flux from
the integrated H$\alpha$ luminosity of the LMC: $Q=3.1 \times 10^{52}\> 
{\rm photons\> s^{-1}}$. This value should be regarded as a lower limit
owing to leakage of some Lyman continuum photons from \ion{H}{2} regions.
Oey \& Kennicutt (1998) estimate
the leakage fraction ranges from $\sim 0\%$ to $\sim 50\%$ for a sample
of 12 bright LMC \ion{H}{2} regions; we adopt their median value of 25\% to
correct the Kennicutt \& Hodge estimate. From the integrated radio continuum
flux, Israel (1980) derives a value for $Q$ of $6.6 \times 10^{52}\>
{\rm photons\> s^{-1}}$, which should be regarded as an upper limit owing
to the nonthermal continuum radiation which has not been subtracted.
We adopt $Q=4 \times
10^{52}\> {\rm photons\> s^{-1}}$ as a conservative estimate. For a mean
metallicity of one-third solar, the measured upper IMF slope of $\sim2.5$
(Massey et al.\ 1995), and an upper mass cutoff of 80 $M_{\odot}$ (both
of which will be assumed throughout), Vacca's (1994)
tables provide an estimate for the total number of O stars in the Large
Cloud of 5530 with an estimated uncertainty of $\sim 50\%$. Thus, the
specific 2--10 keV X-ray luminosity is $5.4 \times 10^{34}\> 
{\rm ergs\> s^{-1}\> star^{-1}}$, a factor of 1.5 to 3 greater than
those derived for the solar neighborhood and the Galaxy as a whole.

\subsection {The SMC}

SMC X-1 is the most luminous HMXB in the Local Group. As discussed in detail
in \S~2, however, its mean observed luminosity is not quite as extraordinary
as is often implied. To complement all of the long-term studies cited above,
we have used the {\sl RXTE\/} ASM database to calculate its mean flux
over the past 4.5 years as we have for the LMC and Galactic binaries. Using
the spectral parameters quoted in \S 2 and a distance of 65 kpc, we find
$L_{\rm x}\> (2-10\> {\rm kev}) = 2.4\times 10^{38}\> {\rm ergs\> s^{-1}}$,
completely consistent with the value found above from monitoring studies
over the past three decades.

Early studies of SMC X-1 with SAS-3 (Clark et al.\ 1978) also led to the discovery of two
other putative HMXBs in the SMC at flux levels only a factor of $\sim5$ lower. One
of these, SMC X-3, has never been seen again, despite sensitive searches by
imaging missions which reached flux levels nearly $10^4$ times
lower. The other source, SMC X-2 which had also disappeared a few months after
its discovery (Clark, Li, \& van Paradijs 1979) has been detected once more in the
intervening 23 years by {\sl ROSAT\/} at a level of $L_{\rm x}\> (2-10\>
{\rm keV}) = 2\times 10^{37}\> {\rm ergs\> s^{-1}}$ (adopting the spectral
parameters of SMC X-1), although a subsequent observation with the same
instrument failed to detect it at a level 650 times lower (Kahabka \&
Pietsch 1996). Both sources have been monitored for the past 4.5 years by
the {\sl RXTE\/} ASM and have been detected with 4~$\sigma$ significance
on only one  and three days, respectively (roughly consistent with the number
of such detections expected by chance, especially considering that the crowded 
region in which they reside raises the systematic uncertainties in daily flux
determinations). Their mean values are both consistent with zero, with
 $2\sigma$ upper limits of 0.05 ASM ct ${\rm s^{-1}}$ or luminosities of
$<8\times 10^{36}\> {\rm ergs\> s^{-1}}$.

As with the LMC, a number of surveys and targeted observations with
{\sl Einstein\/} and {\sl ROSAT}, as well as monitoring observations by
{\sl RXTE\/} and {\sl CGRO\/} have revealed several additional HMXBs and
HMXB candidates in the Small Cloud. One new source, XTE J0111.2-7317 has
had a mean ASM flux of 0.4 ct ${\rm s^{-1}}$ over the last 2 years,
contributing a luminosity of $6\times 10^{37}\> {\rm ergs\> s^{-1}}$
during this interval. However, the source was not detected in either
the {\sl Einstein\/} or {\sl ROSAT\/} surveys of the Cloud at flux levels
more than 100 times lower, so this is unlikely to represent an accurate
estimate of its long-term mean luminosity.  The other two HMXBs included
in the {\sl RXTE\/} monitoring together contribute less than 20\% of this
luminosity, and the remaining thirteen candidates reported in the
literature all have mean fluxes far below this level. Finally, as with
the LMC, the total number of remaining unidentified Cloud members from the
{\sl Einstein\/} (Seward \& Mitchell 1981; Wang \& Wu 1992) and {\sl ROSAT\/}
(Kahabka \& Pietsch 1996; Haberl et al.\ 2000) surveys would, if identified
as HMXBs, increase the integrated luminosity of the population by only
a few percent. Thus, we estimate the total accretion luminosity of the
OB population of the SMC to be $\sim2.7\times 10^{38}\> {\rm ergs\> s^{-1}}$,
or roughly equal to that for the LMC.

It is important to note, however, that more than two-thirds of this luminosity
arises in the singular system SMC X-1 which, in addition to being the most
luminous persistent HMXB in the Local Group, also contains the most rapidly
rotating X-ray pulsar ($P$ = 0.71~s), an object with a spin-up timescale of
only 2000 years. In their detailed study of the spin and orbital evolution
of SMC X-1, Levine et al.\ (1993) estimate that the current high-luminosity
phase of the binary's evolution will last at most a few times the pulsar
spin-up time, or $<5-10\times 10^3$ yr. Compared with the $\sim10^7$ yr
main sequence lifetime of this 20 $M_{\odot}$ star, we have a chance of
$<10^{-3}$ of seeing the system at this X-ray luminosity. Since only
$\sim4\%$ of massive stars end up as short-period HMXBs (Portegies-Zwart
\& van den Heuvel 1999), the number of expected systems in the SMC is
$<0.04\times 10^{-3} \times N_{\rm OB} \sim0.05$. Thus, while it is
not enormously improbable that we see SMC X-1 at this luminosity, the
long-term integrated X-ray luminosity of this galaxy is likely to be
overestimated by a factor of several as a consequence of this one source's
current strut upon the stage. We pursue this matter further below in discussing
the putative dependence of a galaxy's X-ray luminosity on metallicity.

We can estimate the total O star population for the SMC in a manner exactly
analogous to that used for the LMC. Kennicutt \& Hodge (1986) report
$Q=0.8 \times 10^{52}$ photons s$^{-1}$ from H$\alpha$ data, while
Israel (1980) derives $Q=1.3 \times 10^{52}\> {\rm photons\> s^{-1}}$
from the radio continuum emission; using the same considerations cited in the
previous section, we adopt $Q=1\times 10^{52}$ photons
s$^{-1}$. For a metallicity of 0.1 solar, the tables in Vacca yield an estimate of 1300 O
stars. The resulting specific luminosity, then, is $\sim 21 \times
10^{34}\> {\rm ergs\> s^{-1}\> star^{-1}}$ at the present time, although
given the lifetime of SMC X-1 and the arguments presented above, it is likely
to be lower by a factor of $\sim3$ on long timescales, making it more
similar to, but still significantly in excess of, the values derived for the 
solar neighborhood, the Galaxy, and the LMC.

\subsection {M33}

Given its distance (720 kpc), individual X-ray binaries in M33 are
not detectable by non-imaging or ASM instruments, leaving the {\sl Einstein\/}
(Long et al.\ 1981; Markert \& Rallis 1983; Trinchieri, Fabbiano, \& Peres
1988) and {\sl ROSAT\/} (Schulman \& Bregman 1995; Long et al.\ 1996) surveys
as our only views of its X-ray source population. The deepest image is
that from the {\sl ROSAT\/} PSPC obtained by Long et al.\ (1996): 50 sources
were detected within $25'$ of the nucleus above a luminosity threshold of
$1.5\times 10^{36}$ ergs~s$^{-1}$ (for our adopted spectral form of a power-law
spectrum with $\Gamma = 1$ and
$N_{\rm H} = 1 \times 10^{21}\> {\rm cm^{-2}}$---but see below).
Five of the sources are identified with foreground stars, one is a background
AGN, and ten are positionally coincident with optically identified supernova
remnants; since the latter sources have mainly soft X-ray spectra, these
associations are thought to be mostly correct. 

Over 60\% of the soft X-ray luminosity of the galaxy comes from a nuclear
source which is unresolved with the {\sl ROSAT\/} HRI (FWHM = $5''$;
Schulman \& Bregman 1995). The origin of this emission is unknown. There
was evidence from the {\sl Einstein\/} data that the source is variable on
timescales of days to months (Markert \& Rallis 1983; Peres et al.\ 1989);
more recently, Dubus et al.\ (1997) claim evidence for a 20\% modulation
with a 106 d periodicity, although their result is not significant at the
3~$\sigma$ level and the periodicity is apparently inconsistent with the
{\sl Einstein\/} measurement (see their Figure 3). The object's high X-ray
luminosity in the soft band of the imaging experiments is in part a
consequence of the source's soft spectrum. The ASCA observations of
Makishima et al.\ (2000) provide the most detailed spectral data in
the harder X-ray band, and
produce a luminosity estimate of $3.8 \times 10^{38}\> {\rm ergs\> s^{-1}}$.
The unusual
stellar content of the M33 nucleus (O'Connell 1983) and the absence of
obvious signs of an AGN at other wavelengths, has led to a variety of
speculative notions concerning the nature of this source: an anomalous AGN,
a single black-hole HMXB, a cluster of HMXBs, intermediate-mass (Her
X-1--type) binaries, LMXBs, and (predictably) a ``new'' type of X-ray
source. While {\it Chandra\/} observations will soon eliminate many of
these options, it is at present unclear whether some or all of this
source's luminosity should be charged to the OB population's accretion
account. We calculate the specific luminosity for M33 both including and
excluding this contribution.

As for the remaining 33 X-ray sources with $L_{\rm x} > 10^{36}\> {\rm ergs\>
 s^{-1}}$, one (the third brightest) is known to be an eclipsing binary
pulsar (Dubus et al.\ 1999).  In the somewhat unlikely event that all 32
remaining sources also are HMXBs, we can estimate the integrated 2--10 keV luminosity
from the {\it EXOSAT} observations reported in Gottwald et al.\ (1987). The
non-imaging ME detector's field of view includes all the X-ray emission
from M33. The ME count rate was $1.0 \pm0.05$ ct s$^{-1}$ in the 1--6 keV band
and, while not a good fit, the spectrum can be characterized for our purposes 
of estimating a 2--10 keV luminosity by their best-fit power law parameters
of $\Gamma\sim 2.5, N_H\sim 4\times 10^{21}$ cm$^{-2}$. We find a total
X-ray luminosity of $7.7 \times 10^{38}\> {\rm ergs\> s^{-1}}$. Some small
fraction of this emission will be contributed by the soft foreground
stars and M33 SNRs, so we adopt a 2--10 keV luminosity of $7 \times 10^{38}\> 
{\rm ergs\> s^{-1}}$ including the nuclear source, and $3 \times 10^{38}\> {\rm ergs\> s^{-1}}$ if it is excluded.

There is substantial disagreement between the estimated thermal radio continuum
fluxes of M33 between Israel (1980) and Berkuijsen (1983). However, more
recent radio results from Buczilowski (1988) and the H$\alpha$ measurements of
Devereux, Duric, and Scowen (1997) agree quite closely with Berkuijsen's
estimate which we adopt here. The implied Lyman continuum flux is, then,
$Q = 2.5\times 10^{52}$ ph s$^{-1}$; for a
metallicity of 1/3 solar, we derive a total O star population of
3460 for M33. This yields a range for the specific X-ray luminosity
of $9 - 20\times 10^{34}$ ergs s$^{-1}$ star$^{-1}$, a value comparable
to that for the SMC if the nuclear emission is included.

\subsection {M31}

As the largest member of the Local Group, M31 has been studied by all the
major X-ray satellite missions. The {\sl Einstein\/} survey (van Speybroeck et al.\ 1979)
revealed a luminous population of LMXBs both in globular clusters and in
the galactic bulge, plus a disk population presumably consisting of HMXBs
and supernova remnants. The recent {\sl ROSAT\/} PSPC survey (Supper et
al.\ 1997) brought the number of detected sources in the vicinity of the galaxy
to nearly 300, lowered
the luminosity threshold to $10^{35}$ ergs s$^{-1}$, and confirmed the general
picture of two source populations outlined above. In addition to these soft X-ray images, {\sl Ginga\/} carried out
a long pointing at the galaxy in the 2--20 keV band of direct interest here.
Makishima et al.\ (1989), fitted the high signal-to-noise integrated spectrum with a composite model to represent the
dominant LMXB and HMXB populations; the derived fluxes were carefully
corrected for collimator response using the distribution of resolved
sources in the {\sl Einstein\/} images. They find an upper limit to the
HMXB contribution, translated to the 2--10 keV band using their spectral
assumptions (the Galactic foreground absorption of $N_{\rm H} = 6 \times
10^{20}$ cm$^{-2}$ and a cutoff power law with
$\Gamma=1.0$ and $E_{\rm c} = 10$ kev) of $7\times 10^{38}$ ergs s$^{-1}$;
they demonstrate that this is consistent with a value derived by summing
the {\sl Einstein\/} sources in its softer bandpass.

Radio-continuum, far-infrared, and H$\alpha$ images indicate that the bulk
of the star-formation in M31 occurs in a thin ring in the galactic disk
$\sim 10$ kpc from the nucleus (e.g., Beck \& Gr\"ave 1982; Devereux et
al.\ 1994; Walterbos \& Braun 1994; Xu \& Helou 1996).  This star-forming
ring is the region in M31 where high-mass binaries might
be expected to reside.  After eliminating X-ray sources identified with
foreground stars, background AGN, and globular clusters, a comparison of
the PSPC source catalog of Supper et al.\ (1997) and the 60~$\mu$m image of
Xu \& Helou (1996) reveals that 20 {\sl ROSAT\/} sources are positionally
coincident with the star-forming ring; four additional sources are located
close to the ring, four more are coincident with 60~$\mu$m--bright features
outside the ring (excluding the bulge, which is not a site of massive star
formation; see Devereux et al.\ 1994), and six others are found in an outer
spiral arm northeast of the ring where there is some low-surface brightness
IR emission.  Only two of these 34 sources (both of which are weak and in
the ring) are identified with supernova remnants, so the remainder could
all be HMXBs. The PSPC count rates of the HMXB candidates sum to 0.36
ct~s$^{-1}$, roughly 30\% of the 0.1--2.4 keV flux associated with the galaxy.
Applying the Makishima et al.\ spectral parameters to this count rate and
extrapolating to the 2--10 keV band suggests a maximum HMXB luminosity of
$1.5 \times 10^{39}$ in M31. Assuming a somewhat softer spectrum with $\Gamma\sim1.5$ or assigning only half the sources to the HMXB population
yields a result consistent with the {\sl Ginga} analysis: $L_{\rm x}\sim 7.5\times 10^{38}$
erg s$^{-1}$.

As with the other Local Group galaxies, the ionizing photon luminosity of
M31 can be inferred from the thermal fraction of its radio continuum
emission and its H$\alpha$ luminosity.  Beck \& Gr\"ave (1982) estimate
that within the central 20 kpc, the thermal radio flux density at 2.7 GHz
is $\sim 0.68$ Jy, which suggests $Q = 3.2 \times 10^{52}$ photons
s$^{-1}$.  The extinction-corrected H$\alpha$ luminosity of $4.1 \times
10^{40}$ ergs s$^{-1}$ measured by Walterbos \& Braun (1994) gives, assuming
Case B recombination, $n_e = 10^2 - 10^4$ cm$^{-3}$, and $T_e = 10^4$ K,
a nearly identical value for $Q$.  Thus, adopting solar metallicity, we estimate that there are 3660 O stars in M31.  An estimated
luminosity of $7 \times 10^{38}$ ergs s$^{-1}$ for the
HMXB population yields a specific luminosity of $19 \times 10^{34}$ ergs
s$^{-1}$ star$^{-1}$, very similar to the value we obtained for the SMC
and nearly a factor of ten larger than that for the Galaxy. While surprising,
we can think of no plausible loopholes in our argument to eliminate this
difference (see \S 7).

\subsection{Pop I X-rays in the Local Group}

In Table 2, we summarize data relevant to the O-star populations and X-ray
emissivity of the Local Group galaxies. For each object, we give our adopted
distance (uncertain by less than 10\%), the blue luminosity $L_B$ and the
adopted metallicity. The next column lists the quantity $\eta_0$ from
Vacca (1994), the ratio of the number of equivalent O7 stars needed
to produce the observed Lyman continuum flux to the total number of actual
O stars in the galaxy. This quantity depends on metallicity, and on the assumed
slope and mass cutoff of the upper part of the IMF; we have adopted the Salpeter
$\alpha = 2.35$ for the Milky Way and $\alpha = 2.5$ for the other galaxies.
Varying the IMF slope from 2.0 to 3.0 (e.g., Hill, Madore,\& Freedman 1994) changes the O-star counts by --22\% to +60\% for solar metallicity, and 
--22\% to +36\% for a metallicity of 0.1 solar. Likewise, changing $M_{upper}$ 
from $60 M_{\odot}$ to $100 M_{\odot}$ produces changes in the estimated O-star
population of roughly $\pm30\%$. Thus, the IMF parameters are not a major source of uncertainty in our estimates. 

The number of Lyman continuum photons inferred from the observations
described in the text, and the resulting number of O stars are found
in columns 6 and 7. We then include several quantities which depend on the
massive star population: star formation rate, core-collapse supernova rate,
Lyman continuum flux and number of O-stars, all normalized to the value for
the Milky Way. Clearly these quantities are not all independent, but
they are listed to demonstrate that, within a factor of two, these
four quantities are consistent for each galaxy, giving us some confidence that
the O-star numbers by which we normalize our specific X-ray luminosities
are not in error by more than a factor of two. The final two columns
contain our estimates for the 2--10 keV $L_{\rm x}$ discussed above and the X-ray
luminosities per O star which constitute our principal result.

The range of specific luminosities spans an order of magnitude. In
two cases, we list a range of values based on differing assumptions about the
assignment of X-ray flux to Pop I binaries: for M33, we quote the value
including and excluding the nuclear source, and for the SMC, we include the 
current value, as well as one-third of that value based on our arguments about
the lifetime of SMC X-1. In both cases, however, the entire range of allowed
values falls within the extremes defined by the Milky Way and M31. While it
is somewhat curious that the galaxy for which we have the best
information---the Milky Way---has the lowest value, the consistency of the
results we
obtain for the Solar Neighborhood and the Galaxy as a whole adds to the
robustness of this conclusion. Bringing the value for the earlier type galaxy
M31 down by a factor of $\sim 10$ to agree with M0 appears to lie outside of
the range of the uncertainties involved. The implications of this result for
the putative dependence of Pop I X-ray luminosity on metallicity is
discussed below.

\subsection {A Dependence on Metallicity?}

Shortly after the discovery of highly luminous Pop I X-ray binaries in the
Magellanic Clouds, Clark et al.\ (1978) discussed the apparent shift in
the mean X-ray luminosity of HMXBs in the Clouds with respect to that in
the Milky Way, and attributed the higher luminosities of the Cloud binaries
to the lower metallicity of the accreting gas. The discovery that the
metal-poor extragalactic \ion{H}{2} region NGC 5408 has a very high X-ray
luminosity  (Stewart et al.\ 1982) reinforced the notion that the Pop I
X-ray luminosity of a galaxy and its metallicity are inversely correlated.
Alcock \& Paczynski (1978) calculated evolutionary tracks for low-metallicity
massive stars, and pointed out that such stars spend more time in
evolutionary phases with massive stellar winds that power much HMXB emission,
offering a possible explanation for this trend. Hutchings (1984) offered
an alternative explanation, postulating that the fraction of compact objects
in HMXBs that are black holes may be higher in late-type (lower metallicity)
galaxies; indeed, two of the three persistently bright LMC Pop I binaries
are among the best black hole candidates. Without any quantitative analysis
of its significance or cause, numerous studies on the contribution of
starbursts to the X-ray Background (XRB; e.g., Bookbinder et al.\ 1980;
Griffiths \& Padovani 1990) have adopted this $L_{\rm x} - Z$ relation.

The results presented here, however, suggest caution. While our value for the
specific X-ray luminosity per O star in the solar neighborhood is similar to the
number of $1.3 \times 10^{34}$ ergs s$^{-1}$ per O star quoted by Stewart et
al.\ (1982), our values for the LMC and SMC disagree by factors of 4 to 8; a
similar table from Bookbinder et al.\ (1980) contains values higher by yet
another factor of 4. Since no details on the derivation of these numbers
are given in this earlier work, it is difficult to pinpoint the 
causes of these discrepancies,
although the common overestimation of the X-ray luminosities of specific
sources, exemplified by our discussion in \S~2, is a likely culprit. Our
use of the ASM data to obtain long-term mean $L_{\rm x}$ values and our
detailed analysis of the imaging data for each galaxy (as well as modern
estimates for O star counts) has, we hope, reduced the uncertainties in
these estimates.

Our conclusion that M31, the most metal-rich member of the Local Group,
has a specific X-ray luminosity per O star very similar to that of the SMC
(the lowest metallicity galaxy) casts serious doubt on the widely adopted
notion that these two quantities are anticorrelated. The recognition that
SMC X-1, which dominates the value for the Small Cloud, may be sufficiently 
short-lived that the current luminosity of that galaxy is several times
greater than the
long-term average would actually reverse the trend. The detailed census of
2--10 keV point sources in M31, the resolution of the nature of the M33
nuclear source, and the resolution of the point source populations in more
distant galaxies with {\it Chandra\/} should help to constrain further the
values derived here and to clarify the dependence, if any, of X-ray luminosity
on metallicity.

\section {Other X-ray Contributions Associated with Star Formation}

While HMXBs are the most luminous individual X-ray sources arising from
star formation, several other high energy phenomena associated with massive
stars also produce hard
X-rays. For completeness, we evaluate their contributions to the specific
X-ray luminosity per O star here. 

Since most of these phenomena are, like
the HMXBs, associated with all stars down to 8 $M_{\odot}$ (the approximate
dividing line between stars which end their lives in core-collapse supernovae
and those which end as white dwarfs), we include the integrated contributions
from stars down to this mass cut. Furthermore, since these phenomena are mostly
short-lived compared to the main sequence lifetimes of OB stars, we calculate
the expected contribution to the instantaneous X-ray luminosity of the
population by dividing total X-ray luminosity produced by the mean main-sequence
lifetime of the population, weighted by the initial mass function:
   
                $<t_{OB}> = \int\Phi (m) t(m) dm / \int\Phi(m) dm$

\noindent where $\Phi (m) =  Cm^{\alpha}$ is the initial mass function (and we adopt the Salpeter
slope of $\alpha = -2.35$), and $t(m) = 30 (m/10\> M_{\odot})^{-1.6}$
(Stothers 1972). The result, adopting lower- and upper-mass limits of 8 $M_{\odot}$
and 80 $M_{\odot}$, respectively, is $<t_{OB}>\> = 21$ Myr.
 
In the steady state, such as obtains today in the Milky Way, this provides
the appropriate
comparison to our empirical specific luminosity per O star from the HMXBs.
In a galaxy undergoing a starburst with a duration comparable to this timescale,
the relative contributions of these various additional sources of X-ray emission
will be a function of the starburst age. However, for a population of such
starbursting systems, the steady state value provides a valid approximation.

\subsection {Main Sequence O Stars}

As noted in the Introduction, OB stars on the main sequence produce X-rays
which are thought to originate from shocks that develop in unsteady
wind outflows (Lucy \& White 1980; Cooper \& Owocki 1994; Feldmeier et al.\ 1997). The typical ratio of
$L_{\rm x}/L_{\rm bol} \sim 10^{-7}$ (Pallavicini et al.\ 1981). The characteristic temperature
of the emission is $\sim 0.5$ keV (Chlebowski, Harnden, \& Sciortino 1989),
implying a 2--10 keV luminosity of
$\sim 10^{30}$ ergs s$^{-1}$ for all spectral classes. The total is, then, 
$<0.01\%$ of the binary contribution and can be safely ignored.

Harder emission, both thermal and nonthermal, can arise when winds from
neighboring stars collide (Cooke, Fabian, \& Pringle 1978; Chen \& White 1991; Wills, Schild, \& Stevens 
1995). In the Orion Trapezium region, the total
2--10 keV X-ray luminosity, not all of which can reasonably be associated with
this phenomenon is
$ 1.4 \times 10^{33}$ ergs s$^{-1}$  in the 2--10 keV band (Yamauchi \& Koyama
1993; Yamauchi et al.\ 1996). With at least several O stars participating, this
yields a specific luminosity of $<3\%$ that of binary systems. While it is possible that in the
massive OB associations found in starburst nuclei wind collisions could
be significantly enhanced, it seems highly improbable that they will
compete with binaries as a significant source of an OB star population's
hard X-ray luminosity.

\subsection {Pre-Main Sequence Stars}

Lower mass stars, formed in association with massive stars, undergo a T Tauri 
phase prior to descending onto the main sequence during which significant 
hard X-ray emission is produced (e.g., Koyama et al.\ 1996). Again, using the local example of Orion as a template, the X-ray luminosity associated 
with T Tauri stars in the 2--10 keV band is
$1.8\times 10^{33}$ ergs s$^{-1}$ (Yamauchi \& Koyama 1993). Since there are $\sim50$ O stars in the Orion complex, this yields a specific luminosity
of $\sim 4\times 10^{31}$ ergs s$^{-1}$. 
If there is a discrepancy in the ratio of high- to low-mass
stars in starburst galaxies versus the local sites of star formation, it
is likely to be
in the direction of a deficit of lower mass stars, reducing this contribution
to an even smaller value. In any event, it appears unlikely that the
contribution from pre-main sequence low-mass stars will exceed 1\% that
of the HMXBs.

\subsection {Supernova Remnants}

The violent deaths of massive stars in core-collapse supernovae provide
several means of producing X-ray emission: thermal emission from shock-heated
gas left by the passage of the SN blast wave, nonthermal emission from
particles accelerated at the shock front, nonthermal emission from
a synchrotron nebula generated by a young, rapidly rotating neutron star,
and emission from a hot young neutron star's surface and magnetosphere. We
examine each of these in turn, taking the Galaxy's supernova remnant (SNR) 
population as exemplary.

The hot gas generated by the outward moving shock wave from the SN explosion,
along with the stellar ejecta heated by the reverse shock, produce thermal
X-ray emission with a temperature characteristic of the shock velocities;
for most of a remnant's life these range from 300 to 3000 km s$^{-1}$,
yielding nominal temperatures from 0.2 to 20 keV, although delayed
equilibration between the protons and electrons, nonequilibrium ionization,
and inhomogeneities in the ambient and ejected material conspire to produce
observed temperatures for the bulk of the emitting material of $\sim 1$ keV.

While more sophisticated models of remnant X-ray emission have been constructed
over the past few decades, it suffices for our purposes of estimating the
total 2--10 keV energy radiated to use the simple Sedov equations (e.g.,
Gorenstein \& Tucker 1976). For typical SNR parameters (explosion energy $E_0 
\sim 10^{51}$ ergs, ambient density $n_0\sim 1$ cm$^{-3}$), we have calculated
the temperature, shock velocity, and radius, as well as the
fraction of the radiated flux emitted in the 2--10 keV band, as a function of 
time. As the swept-up material decelerates the shock, the temperature falls and
the X-ray luminosity rises. However, the fraction of the emission in the 2--10
keV band also falls once $v_s < 1300$ km s$^{-1}$ ($t\sim 1500$ yr), such
that, for $10^3 {\rm yr}<t<10^4$ yr, the 2--10 keV band luminosity
is constant to within a factor of two, with an average value of $7\times10^{35}$ erg s$^{-1}$; for later times, the emission
in this band rapidily declines into insignificance. Thus, the integrated
contribution from thermal remnant emission is $\sim 7\times10^{35}$ erg s$^{-1}
\times 10^4 {\rm yr} = 2.2\times 10^{47}$ erg; dividing by our mean O-star
lifetime $t_{OB}$ gives $3.3\times 10^{32}$ erg s$^{-1}$ per O star or roughly 1-2\% of
the HMXB contribution.

In addition to heating ambient gas and supernova ejecta, the shock wave
sweeps up magnetic fields and accelerates particles to relativistic
energies. The primary consequence of this is the bright radio emission
associated with SNRs. However, for young remnants at least,
the particle spectrum extends to very high energies, producing detectable
synchrotron radiation in the X-ray band. The 2--10 keV X-ray luminosity
of the historical remnant SN1006 is dominated by such synchrotron emission
(Koyama et al.\ 1995), and evidence for such nonthermal radiation has recently
been detected in several other young remnants (Petre et al.\ 
1999 and references therein). Indeed, Petre et al. claim that there is 
evidence that {\it all} young remnants have an X-ray synchrotron component,
and that we only see this as a dominant contributor to the remnant's X-ray 
emission when the SN takes place in a very low density region of the
interstellar medium and thus can form no significant reverse shock to
illuminate the ejecta. The synchrotron luminosities of these sources are
typically $<10\%$ of the thermal $L_{\rm x}$, and the timescale over which 
this component is significant is less than that for the thermal emission. Thus,
its overall contribution to the hard X-ray luminosity is almost certainly
$<1\%$ that of the HMXB contribution.

One of the most luminous hard X-ray sources in the Galaxy is the Crab Nebula,
a remnant of the supernova of 1054 AD powered by rotational kinetic energy
loss from the young neutron star created in the explosion; in the 2--10 keV
band, $L_{\rm x} = 1.5 \times 10^{37}$ ergs s$^{-1}$ (Harnden \& Seward 1984). While often
characterized as the prototypical young neutron star, the Crab is, in fact,
not typical.  For example, the SN of 1181 AD also produced a pulsar-powered
synchrotron nebula
(3C 58), but its 2--10 keV X-ray luminosity is $10^3$ times lower at only $1.4 \times 10^{34}$ ergs
s$^{-1}$, despite its slightly younger age (Helfand, Becker, \& White 1995). Furthermore, evidence for
young pulsars in the remnants of other core collapse supernovae has been
notoriously difficult to find, and while more than three dozen such cases of
SNR/neutron star
associations have now been suggested, none produce X-ray luminosities
within a factor of five of the Crab pulsar (see Helfand 1998
for a review). Broad distributions of initial spin period and magnetic field
strength for newly born neutron stars are likely to be responsible for the
wide range of properties observed.

A firm upper limit on the contribution such objects can make to the hard X-ray
luminosity of a young stellar population can be derived by assuming that all
neutron stars are born with $P\sim 10$ msec, yielding a total rotational
kinetic energy of $E_{\rm rot} = {1\over2} I \omega ^2 \sim 2 \times 10^{50}$ erg,
where $I~(\sim 10^{45}$ g cm$^2$) is the star's moment of inertia and $\omega$ is the rotational
frequency. For the Crab, the fraction of the rotational kinetic energy loss
rate $\dot E$ emerging in the 2--10 keV band is $\eta_{\rm x} = 0.05$; other young
Crab-like remnants such as 0540-693 in the LMC and 1509-58 show similar
ratios of $L_{\rm x}/ \dot E$. Thus, the upper limit to the contribution of young
pulsar nebulae to the 2--10 keV luminosity of an OB population is $\sim
2 \times 10^{50}$ erg $\eta_{\rm x} f_{ns} f_{Lc} / t_{OB}$, where $f_{ns}$ is
the fraction of supernovae that produce neutron stars, and $f_{Lc}$ is the
fraction of the Crab spin-down luminosity of the average young neutron
star. Although the mass cut dividing black hole and neutron star remnants
of core collapse events is unknown, $f_{ns}$ is likely to be of order
unity. The quantity $f_{Lc}$ is less well-determined, but is clearly much
less than unity: for a core-collapse SN rate of one per century, there
should be ten sources with $L_{\rm x} > L_{\rm Crab}$ in the Galaxy.
In fact, there is only one source at 0.2 $L_{\rm Crab}$ (G29.7-0.3; Helfand \&
Blanton 1996) and no other sources within an order of magnitude. We adopt
$f_{Lc}=0.2$, although we regard this as a conservative upper limit. Thus,
the X-ray luminosity contribution from pulsar synchrotron nebulae could be
as high as $\sim 3 \times 10^{33}$ ergs s$^{-1}$ per O star or roughly 10\% of
the HMXB contribution; if, as seems to be the case in the Galaxy, the median
X-ray luminosity of young neutron stars is at least a factor of ten less than that of the Crab, the contribution of synchrotron nebulae will be $<5\%$ of the
HMXB value.

The final source of X-ray emission resulting from a SN explosion is the
thermal emission from the hot surface of the young neutron star and the
nonthermal emission from its magnetosphere. Since rapid neutrino cooling
reduces the surface temperature to under $5 \times 10^6$ K within a few decades,
the contribution of thermal emission in the 2--10 keV band is completely
negligible.  Nonthermal pulsed emission in the Crab accounts for only
$\sim3\%$ of the total nonthermal emission produced by the pulsar nebula.
Becker and Trumper (1997) have shown that $L_{\rm x} /\dot E \sim 10^{-3}$ for a wide range of pulsar ages and magnetic field strengths; thus, magnetospheric X-ray
emission from rotation-powered pulsars is negligible compared to the HMXB
contribution.

\subsection {Galactic Winds}

The large mechanical energy input to the interstellar medium of a starburst
galaxy from stellar winds and supernovae results in a pressure-driven wind
of hot plasma. Such ``superwinds" (e.g., Heckman, Armus, \& Miley 1990) have been observed to be characteristic
of galaxies with high star formation rates, and diffuse X-ray emission
associated with them has been detected in a number of galaxies (e.g.,
M82, NGC 253 [Fabbiano 1988]; NGC 3256 [Moran, Lehnert, and Helfand 1999],
etc.). The
characteristic temperature of these winds, however is $\sim 0.3$ to $\sim 0.8$
keV, and their contribution to the galaxies' emission above 2 keV is
negligible.

\section {HMXBs and the Hard X-ray Luminosities of Starburst Galaxies}

Having characterized the dominant role of HMXBs in the production of hard X-rays
in the Milky Way and other Local Group galaxies, we can now discuss
the direct contribution of OB stars and their remnants to the total hard X-ray 
luminosities of galaxies undergoing bursts of star formation.  To do this, we 
need to relate the specific X-ray luminosity per O star to observable quantities
for nearby starbursts---total hard X-ray flux and infrared luminosity.

The integrated 2--10 keV X-ray luminosity of high-mass binaries in a
star-forming galaxy can be expressed as $L_{_{\rm X}} =
[L_{_{\rm X}}/N({\rm O})]_{_{\rm HMXB}} \times N({\rm O})$, where
$[L_{_{\rm X}}/N({\rm O})]_{_{\rm HMXB}}$ is an adopted value of the
specific X-ray luminosity per O star for HMXBs, and $N({\rm O})$ is
the actual number of O stars present.  Assuming an IMF slope of 2.35,
an upper mass cutoff of 100 $M_{\odot}$, and solar metallicity, the
models of Leitherer \& Heckman (1995) predict that a region producing
stars at a constant rate of 1 $M_{\odot}$ yr$^{-1}$ for at least $10^7$ yr
will have $2.5 \times 10^4$ O stars and an associated bolometric luminosity
of $5.3 \times 10^{43}$ ergs s$^{-1}$.  Provided that the young stellar
population dominates the host galaxy's bolometric luminosity---which is
approximately equal to its total infrared luminosity $L_{_{\rm IR}}$---the
number of O stars can be scaled for a system of arbitrary star-formation
rate: $N({\rm O}) = 2.5 \times 10^4\> (L_{_{\rm IR}} / 5.3 \times 10^{43})$.
The binary luminosity expression then becomes
$L_{_{\rm X}} = 4.7 \times 10^{-40}\>
[L_{_{\rm X}}/N({\rm O})]_{_{\rm HMXB}}\> L_{_{\rm IR}}$,
or in terms of fluxes, $F_{_{\rm X}} = 4.7 \times 10^{-40}\>
[L_{_{\rm X}}/N({\rm O})]_{_{\rm HMXB}}\> F_{_{\rm IR}}$.

Using the latter equation, we have computed the range of HMXB X-ray fluxes
expected at a given IR flux for the range of Local Group values of
$[L_{_{\rm X}}/N({\rm O})]_{_{\rm HMXB}}$.  These are represented by the shaded
region in Figure 1.
This region is bounded on the lower-right by the specific X-ray
luminosity per O star derived from direct counts of HMXBs and O stars in
the solar neighborhood ($1.6 \times 10^{34}$ ergs s$^{-1}$ star$^{-1}$),
and on the upper-left by the value of $20 \times 10^{34}$ ergs s$^{-1}$
star$^{-1}$ obtained for the SMC and M31.  The dashed line represents the upper
limit derived for the solar neighborhood which is roughly equal to the
global Milky Way value.  Also plotted in Figure 1 are
the locations in the $F_{_{\rm X}} - F_{_{\rm IR}}$ plane of several nearby
starburst galaxies that have been studied with {\sl ASCA}.  The IR fluxes
of these objects have been calculated from the highest reported {\sl IRAS\/}
flux densities using the $F_{_{\rm IR}}$ prescription of Sanders \&
Mirabel (1996).  Their 2--10
keV X-ray fluxes have been collected from published {\sl ASCA\/} results
(references are provided in the figure caption).  Note that the X-ray luminosities of
the starbursts span several orders of magnitude, from
$< 10^{39}$ ergs s$^{-1}$ (NGC 1569 and NGC 4449) to $\sim 10^{40}$
ergs s$^{-1}$ (NGC 253 and NGC 2146) to $> 10^{41}$ ergs s$^{-1}$
(NGC 3256 and NGC 3690).

Several important conclusions can be drawn from Figure 1.  First, there is
a clear tendency for the 2--10 keV X-ray fluxes of starburst galaxies to
increase with $F_{_{\rm IR}}$, indicating that their hard X-ray luminosities
are largely governed by sources whose contributions are proportional to
the star-formation rate.  As discussed in the previous section, HMXBs are
expected to dominate over all other such contributors.  However, in order for
HMXBs to account for {\it all} of the hard X-rays produced in starbursts,
their typical output per O star must be significantly greater than that
observed in the Milky Way or the LMC.  Even in the starburst galaxies
with the lowest $F_{_{\rm X}}/F_{_{\rm IR}}$ ratios (NGC 1569, NGC 3256,
M83, and NGC 253), HMXBs would have to exhibit $[L_{_{\rm X}}/N({\rm O})]_{_{\rm HMXB}}$
values that are 5 times higher than the Milky Way's.  Both direct observations
and population syntheses
(e.g., Dalton \& Sarazin 1995) indicate that the bulk of the X-ray emission
of a binary population arises from the small fraction of objects with
the highest individual luminosities.  Thus, if HMXBs produce most of the
hard X-ray flux of starburst galaxies, we would expect such systems to
have many more high-luminosity objects (per O star) than the Milky Way
and the LMC.  For the nearest starbursts (including several of the objects
in Fig.~1), this hypothesis is testable with high-resolution
{\it Chandra\/} observations.

The good correlation between $F_X$ and $F_{\rm IR}$ stands in contrast to the large
scatter in a plot of $F_{\rm X}$ vs.\ $F_B$, the blue optical flux, from these same galaxies.
Thus, unlike normal galaxies which show a tight correlation between $F_{\rm X}$ and
$F{\rm _B}$ (Fabbiano 1989)---i.e., the X-ray luminosity is proportional to the
light from the whole stellar population and is dominated by long-lived, low-mass
X-ray binary emission---in starbursts, the dominant X-ray production is
associated with the young stellar population.

Two objects, M82 and NGC 3310, deviate significantly from the $F_{_{\rm X}}
- F_{_{\rm IR}}$ trend exhibited by the other starburst galaxies in
Figure 1.\footnote{M82 contains a variable hard X-ray source (Matsumoto
\& Tsuru 1999) that can, in outburst, dominate the 2--10 keV emission of
the galaxy.  However, preliminary {\it Chandra\/} results (Griffiths et al.\
1999) indicate that there are many other sources of hard X-ray emission
in M82 as well.  The X-ray flux of M82 shown in Figure 1, from Moran \&
Lehnert (1997), is the {\it lowest} flux measured by {\sl ASCA}; it is likely to
represent the integrated contribution of the ``quiescent'' sources of hard
X-ray luminosity in the galaxy.}  The X-ray fluxes of these two objects
are clearly inconsistent with the level of emission expected from an HMXB
population, even one similar to that of the SMC, suggesting that
each galaxy possesses an extra component of hard X-ray luminosity that is
weak (or absent) in the other starbursts.  Indeed, {\it Chandra\/} images
of NGC 253 (Strickland et al.\ 2000) and M82 (Griffiths et al.\ 1999),
which have similar IR luminosities but 2--10 keV X-ray luminosities that
differ by at least a factor of 5, reveal strikingly different hard X-ray
morphologies.  The hard X-ray flux of NGC 253 is produced almost entirely
by discrete sources, whereas in the more luminous M82, about half of the
hard X-ray emission arises from a diffuse component coincident with the
most active region of star formation.  We have suggested
previously (Moran \& Lehnert 1997; Moran, Lehnert, \& Helfand 1999) that
inverse-Compton scattered emission, resulting from the interaction of IR
photons with supernova-generated relativistic electrons, may in some
circumstances contribute appreciably to the hard X-ray fluxes of starburst
galaxies.  M82 and NGC 3310 thus represent the best sites for the
investigation of this possibility.

\section {Starbursts and the X-ray Background}

We have demonstrated that the hard (2--10 keV) X-rays produced directly by a
population of OB stars, their remnants, and their accompanying lower-mass
brethren are dominated by the small fraction of massive stars that form X-ray 
binaries. Despite our efforts to assess with care the long-term mean luminosities
of such systems through our use of the ASM database and our consistent
methods for deducing O-star number counts, we find a range of an order of
magnitude in the specific X-ray luminosity per O star among the galaxies of the
Local Group. While the specific luminosities of M33 and the SMC are each
dominated by a single source, and, as we have argued, could plausibly have a
long-term value within a factor of 2 of the Milky Way, M31 remains an outlier.
It is possible that only a small fraction of the 34 luminous X-ray sources
coincident with star-forming regions in that galaxy are HMXBs, although no other Local Group
galaxy shows a similar population of bright, non-HMXB objects. Alternatively,
the O-star population of M31 could have been severely underestimated if a
large fraction ($>80\%$) of its Lyman continuum photons escape from the
galaxy without ionizing a hydrogen atom. Were either (or both) of these 
scenarios to hold, and were we to ignore the bright single sources in M33 and
the SMC (a somewhat uncomfortable chain of assumptions), the specific X-ray
luminosity per O star in Local Group members could all fall within a factor of 
two of $5\times 10^{34}$ erg s$^{-1}$ per O star.

In this case, the starburst galaxies would all require a source of X-ray
luminosity in addition to the direct contributions of the OB star population.
Even if we allow the full observed range of specific luminosities and brand
the Milky Way as atypical, however, some starbursts still require an additional
hard X-ray component. While a buried active nucleus is a plausible
candidate, {\it Chandra\/} observations have ruled this out in the case of M82.
The diffuse nature of a significant fraction of the hard X-ray flux from M82 is
consistent with our suggestion of IC emission as the origin of this additional
component. The fact that the less intense and more diffuse
starburst in NGC 253 (1) shows no significant diffuse hard emission, and (2) falls
within the $L_{\rm x}/L_{IR}$ band predicted from binaries alone, is also consistent
with this picture, since the predicted IC luminosity from NGC 253 would be
negligible.

Natarajan and Almaini (2000) have recently used global energetics arguments
to conclude that OB stars and their products (HMXBs and SNRs) contribute at most
$\sim 1\%$ of the X-ray background at energies above 2 keV. They (reasonably)
assume the HMXB population tracks the global star formation rate, although their
normalization assumes a Milky Way hard X-ray luminosity a factor of 2.5 lower 
than we derive in \S3.2, and, thus, a factor of $\sim 10$ below the mean value 
for the Local Group. In addition, as they note, extra contributions such as IC 
emission are not included in their calculation. Thus, we conclude it
remains plausible that a significant contribution to the hard X-ray background
arises from starburst galaxies. 

It should be emphasized that such a conclusion is not inconsistent with
existing deep-field X-ray source counts or faint X-ray source identifications.
Given the steep redshift dependence of the star formation rate, the vast
majority of the XRB contribution from starbursts will arise at redshifts
greater than 1. To illustrate, we use the results of Moran et al.\ (1999) in
which we showed that, owing to the tight correlation between far IR and
centimetric radio emission for starburst galaxies (and the correlation shown
in Figure 1 between far-IR and X-ray luminosity in these same galaxies), faint
radio source counts can be used to constrain the surface density of starburst
XRB contributors. For a 75\% starburst fraction (Richards 1998) in the $\sim 
5{\rm mJy} > S_r> 5\mu{\rm Jy}$ radio flux density range, $N(>S)=17.4
S_r^{-1.2}$  arcmin$^{-2}$ (Fomalont et al.\ 1991).  The {\sl ROSAT\/}
Deep Survey in the Lockman 
Hole (Hasinger et al.\ 1998) had a 0.5--2.0 keV limit of $5 \times 10^{-15}$
erg cm$^{-2}$ s$^{-1}$ for their complete sample of 50 sources over $\sim 0.1$ 
deg$^2$. Using the ratio of 5 keV to 5 GHz flux density for starbursts found in
Moran et al.\ (1999), $R_{5,5} \sim 10^{-18}$ ergs cm$^{-2}$ s$^{-1}$ keV$^{-1}
\mu{\rm Jy}^{-1}$, this implies an equivalent radio flux density limit of
$\sim 9$ mJy;\footnote{To compare the radio source counts to the {\it core\/}
X-ray/radio flux density ratio, we multiply the core radio flux density by
a factor of 2 (a typical ratio for local starbursts). Even if we assume the
entire radio flux density is from the nuclear starburst, we still expect
$<0.25$ and $<0.4$ sources in the {\sl ROSAT\/} and {\it Chandra\/} surveys,
respectively.} assuming an X-ray spectral index of $\Gamma=1.7$ as observed
in NGC 3256, we should then expect $\sim 1.2$ starbursts deg$^{-2}$ or 0.1
such sources in the survey. For the largest {\it Chandra\/} Deep Survey published
to date (Giacconi et al.\ 2000), the 2--10 keV limit is $2\times 10^{-15}$ erg
cm$^{-2}$ s$^{-1}$ which corresponds to a 5.7 mJy radio flux density and
an expected surface density of 2.2 sources deg$^{-2}$, or $\sim 0.2$ sources
in  the 0.096 deg$^2$ survey area. 

In summary, the deepest surveys yet performed 
have not gone deep enough to reveal the population of starbursts at their
expected luminosities. The continued flattening of the AGN-dominated
X-ray log$N$-log$S$ seen by {\it Chandra\/} observations at 2--10 keV flux levels above
$10^{-15}$ erg cm$^{-2}$ s$^{-1}$ strengthens the requirement for a new 
population of objects at fainter fluxes in order to account for the
remaining 20-25\% of the X-ray background. Starburst galaxies remain an 
attractive candidate, and deeper {\it Chandra\/} surveys should begin to find
them at a surface density of 30 deg$^{-2}$ ($\sim3$ per {\it Chandra\/} field) when
a flux threshold of $4\times 10^{-16}$ erg cm$^{-2}$ s$^{-1}$ is reached.

\acknowledgments

DJH is grateful for the support of the Raymond and Beverly Sackler Fund,
and joins ECM in thanking the Institute of Astronomy of the University of
Cambridge for hospitality during much of this work. This research has made
use of data obtained from the High Energy Astrophysics Science Archive
Research Center (HEASARC), provided by NASA's Goddard Space Flight Center.
The work of ECM is supported by NASA through {\it Chandra\/} Fellowship
PF8-10004 awarded by the {\it Chandra\/} X-ray Center, which
is operated by the Smithsonian Astrophysical Observatory for NASA under
contract NAS8-39073. DJH acknowledges support from NASA grant NAG 5-6035.
This paper is Contribution Number 696 of the Columbia Astrophysics Laboratory.

{}

\clearpage

\begin{figure}
\epsscale{0.7}
\plotone{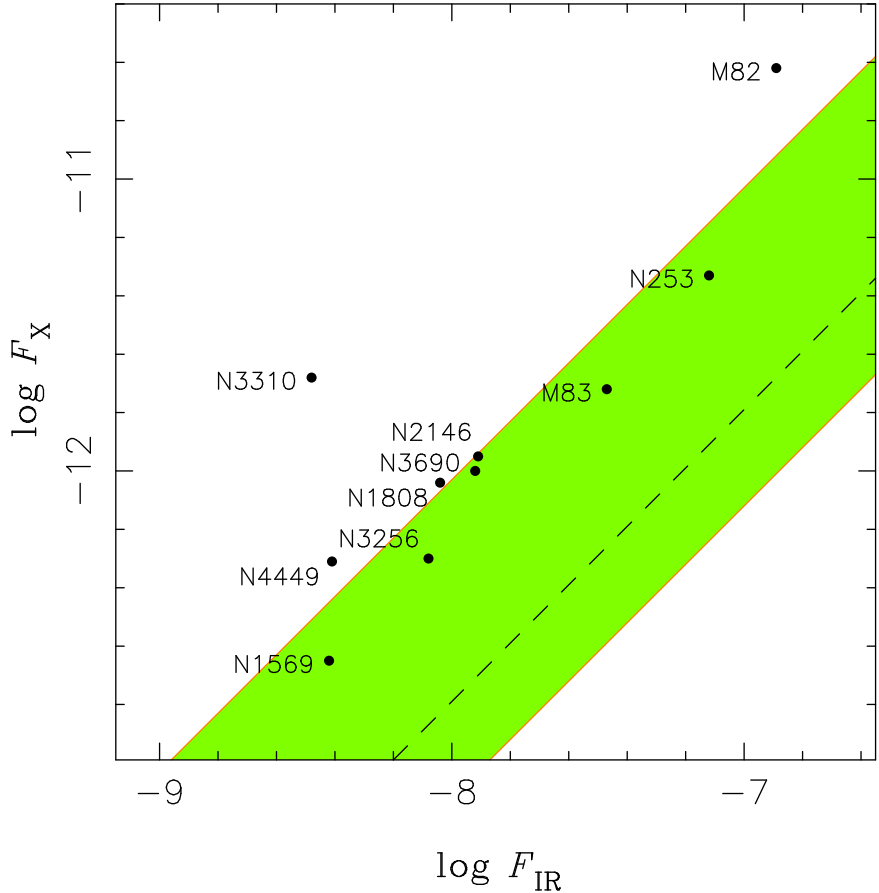}
\caption[]{X-ray flux in the 2--10 keV band versus 8--1000 $\mu$m infrared
flux for ten starburst galaxies that have been studied with {\sl ASCA}.
The infrared fluxes are computed using the highest reported 12, 25, 60,
and 100 $\mu$m {\sl IRAS\/} flux densities and the formula $F_{_{\rm IR}}
= 13.48\> S_{12} + 5.16\> S_{25} + 2.58\> S_{60} + S_{100}$ (Sanders \&
Mirabel 1996).  The X-ray fluxes are from published {\sl ASCA\/} studies
(NGC 253: Ptak et al.\ 1997; NGC 1569: Della Ceca et al.\ 1996; NGC 1808:
Awaki et al.\ 1996; NGC 2146: Della Ceca et al.\ 1999; M82: Moran \&
Lehnert 1997; NGC 3256: Moran, Lehnert, \& Helfand 1999; NGC 3310 and
NGC 3690: Zezas, Georgantopoulos, \& Ward 1998; NGC 4449: Della Ceca et
al.\ 1997; M83: Okada et al.\ 1997).  As described in the text, the shaded
region shows the range of X-ray fluxes expected from HMXBs at a given IR
flux for the different determinations of the specific X-ray luminosity per
O star in Local Group galaxies.  The region is bounded on the lower-right
by the value of $1.6 \times 10^{34}$ ergs s$^{-1}$ star$^{-1}$ found for the
solar neighborhood, and on the upper-left by the SMC value of $20 \times
10^{34}$ ergs s$^{-1}$ star$^{-1}$.  The dashed line represents the upper
limit for the solar neighborhood of $3.5 \times 10^{34}$ ergs s$^{-1}$
star$^{-1}$.  Clearly, if HMXBs are responsible for the bulk of the hard
X-ray emission from starburst galaxies, their output per O star must similar
to that observed in the SMC.}
\end{figure}

\end{document}